\documentclass[12pt]{article}             
\def\={\!=\!}

\usepackage{amsmath,amsthm,amsfonts,latexsym,amscd,amssymb,enumerate}
\usepackage{color}

\newcommand{\red}[1]{\begin{color}{black}{#1}\end{color}}

\newcommand{\black}[1]{\begin{color}{black}{#1}\end{color}}

\def\={\!=\!}

\def\|{{\Vert}}

\def\r{r}
\def\c{}

\def\R{{\Bbb R}}
\def\N{{\Bbb N}}
\def\noi{\noindent}

\newcommand\skipv{\vspace{2mm}}
\newcount\nR
\newcount\nC
\newcount\nT
\newcount\nL
\def\Remark{\vspace{2mm}\noi{{\bf{Remark \number\nR. }}} \advance\nR by 1}
\def\Corollary{\vspace{2mm}\noi{{\bf{Corollary \number\nC. }}} \advance\nC by 1}
\def\Theorem{\vspace{2mm}\noi{{\bf{Theorem \number\nT. }}}\advance\nT by 1}
\def\Lemma{\vspace{2mm}\noi{{\bf{Lemma \number\nL. }}} \advance\nL by 1}
\def\Proposition{\vspace{2mm}\noi{{\bf{Proposition \number\nL. }}} \advance\nL by 1}

 \title{Games of Incomplete Information and Myopic Equilibria}
\author {R. Simon, S. Spie{\.z}, H. Toru{\'n}czyk}

\begin{document}

\maketitle

\thispagestyle{empty}
\vfill

\noi London School of Economics\newline 
Department of Mathematics\newline 
Houghton Street\newline 
 London WC2A 2AE\newline 
 
 \noi Institute of Mathematics\newline
 Polish Academy of Sciences\newline
 {\'S}niadeckich 8, 00--656 Warszawa

\vfill

\date{}
  
\setcounter{page}{-1}
  \noi      

\newpage
\vskip2cm
\thispagestyle{empty}

\noi 
    {\bf Abstract:} \ \  Combine  two games of incomplete information, one  after the other,    for which their  equilibria are established through  very  different   methods; will the resulting composite   game  have an equilibrium?  Let $\Gamma$ be the set of probability distributions  on the pathways of play of the first game and for  every subset $C$ that can be held in common knowledge  at the conclusion of the first game and $\gamma \in \Gamma$  let $\gamma | C$ be  the conditional probability distribution on $C$, given that it is   well defined.   If the first game has a finite  game tree with perfect recall and \red{ for every such $C$ the second game has  an equilibrium payoff correspondence which, as a  function of   $\gamma| C$, is  upper-semi-continuous and  has non-empty, convex and compact values}, then  the answer is yes.    To prove this, the  concept of a   myopic  equilibrium is introduced, an alternative equilibrium concept
    to that of the Nash equilibrium. In spite of the 
     difference, there is
     a strong relationship between the two equilibrium concepts
     in the context of incomplete information and  repetition.

\vskip2cm

\noi {\bf Key words}: Repeated games and game trees, topological structure of  equilibria, fixed points, the nearest point retraction onto a simplex

\newpage

\section{Introduction} 
The following game inspired this work.  Nature chooses  a state $k$ from a finite set $K$ according to some 
 probability distribution $p_0\in \Delta (K)$.   There are two players, Player One and Player Two. Player One, but not Player Two,  is informed of 
 nature's choice.  The players choose actions simultaneously 
 which are commonly 
  observable directly after those choices, and this situation is repeated an infinite 
 number of times, but with nature's choice of $k$
  fixed from the start. The payoffs to both  player are determined 
  both by what the two players do and  by nature's choice. If 
 the  payoffs to the players are determined by 
 the limit behaviour of the average payoffs, 
  a infinitely repeated undiscounted game 
 of incomplete information on one side has been described in Aumann and Maschler (1995),
 and the existence of their  equilibria 
 was established in Simon, Spie{\.z}, and Toru{\'n}czyk (1995).
 We introduce the following game variation. For both players $i=1,2$ there  are 
  finitely many  non-negative values
 $\lambda^i_1, \lambda^i_2, \dots , \lambda^i_n$ 
 with $0< \lambda_1^i+ \dots + \lambda_n ^i =\lambda^i <1$
so that  for Player $i$  the $j$th stage is weighted according to
 $\lambda^i_j$ and the undiscounted infinitely repeated game
 is weighted according 
to $1-\lambda^i$.    Does such a  game have a Nash  equilibrium? 

\vskip.2cm 
\noi The above question was posed to us by A. Neyman (private communication, 2016),
and our initial response to his question was `` definitively not!''
The proof of  equilibrium existence 
 for a game  with a finite tree structure and perfect recall    uses fixed point theory,
 through the original   proof of Nash equilibria in Nash (1950)   and the  
 application  of Kuhn's Theorem (1953).  
However the proof of equilibrium existence for the undiscounted infinitely repeated 
 game uses a covering theorem that has similarity  
 with the Borsuk-Ulam Theorem (and yet  neither implies  nor is 
 implied by the Borsuk-Ulam Theorem). 
 Why should there be a synthesis of these two very different 
  proofs? 
 \vskip.2cm 
\noi We  answer A. Neyman's question in  the affirmative. 
We make no  synthesis of the  two  proofs, rather we 
 apply  properties of the equilibria of  the infinitely repeated games
  to the finite stage game.
To answer this question  
  we introduce a new equilibrium concept, called a {\em myopic}  equilibrium. 
 \vskip.2cm 
\noi What is the main problem with  understanding the 
 strategic aspects of  finitely many initial  stages 
 followed by an  infinite stage game? Lets simplify the problem, 
 so that $\lambda^1_1 = \lambda^2_1= \lambda=\frac 13$,
 meaning that the first stage counts for $\frac 13$ of the payoff and
 all the remaining infinitely many stages count for $\frac 23$ of the payoff.  
 As Player Two knows nothing but $p_0$ about the state of nature on 
 the first stage, he  must choose some state independent 
 mixed strategy $\tau$. As different states 
  could have very different payoff 
 structures, one would expect   large 
   initial payoff  advantages for Player One     through 
  actions  that are dependent on the 
  states of nature. But by  doing so, Player One could  
  reveal too  much about the  state, as 
   $\frac 23$   of the payoff comes 
 from the following stages. Both using her information too much and not at all 
 seem   to be 
  foolish  options  for Player One. 
 There is a delicate give and take between the initial 
 choices  of Player One at the different states and the conditional probabilities on the states that these choices 
 induce. 
   \vskip.2cm
 
 \noi   By a pure  strategy
 of Player One on the first stage we mean a determination of an initial action 
 dependent on the state of nature, so that if $I$ is her set of actions then there are 
 $|I|^{|K|}$ different pure strategies for the first stage.  By a mixed strategy of Player One 
 on the first stage we mean a probability distribution over those 
 pure strategies.      Lets assume that  there is a subset of  
  the equilibrium payoffs 
 of the infinitely repeated game that change continuously with the conditional 
 probability distribution on the states of nature. 
Keeping with the idea that $\lambda=\lambda^i_1$ for both $i=1,2$, we could define a game where 
 on the first stage Player One chooses a mixed strategy for herself,
 Player Two chooses a mixed 
 strategy for himself (necessarily state independent),
 followed by a payoff determined by the initial stage and  
  an  equilibrium payoff associated with
 the induced conditional probability on the states 
 of nature.
 For any fixed mixed strategy of Player Two, the payoff  for Player One  
 will not in general be concave as a function of her mixed strategies.   
 As there are many games that don't have   
 any equilibrium  when a  payoff function is  not concave relative to the  actions of
  the player concerned (see later example),
 we anticipated that the   composite game of Neyman's 
 question  
 would  fail to have an equilibrium. 
\vskip.2cm 
\noi On closer examination, we discovered that replacing 
 a  mixed strategy of Player One with another mixed strategy of the same 
 player    
 was not the correct model of strategic deviation for these games. Assume that  Player One 
 is committed to  some mixed strategy on the first stage 
  for which every  action 
  is played with large positive probability at some state of nature.
  No matter what Player One does on that first stage there is 
 no  appearance of a strategic 
 deviation. Unless Player One demonstrates an action that should never have happened, 
  Player Two will continue to interpret the future  actions 
 of Player One according to a putative  commitment to that  mixed 
 strategy, and not to a different strategy that Player One might have chosen.
 For there to be an equilibrium it is necessary 
 that   any advantage 
  from one  action    
  is properly offset by a subsequent disadvantage on 
 the following stages, and that this holds simultaneously for all states of nature.   
 We will see that the required equilibrium 
  property  leads directly to the  definition  of 
 a myopic equilibrium. 
 \vskip.2cm 
\noi The myopic equilibrium concept was formulated to solve Neyman's problem, 
 as explained above.  Its main   application in this paper   
 is however broader,   to game trees  
   where  information is  incomplete. It is intriguing that in order 
 to prove that a certain game has a Nash equilibrium, it was necessary 
 to formulate a new equilibrium concept that can differ greatly  from 
 the Nash equilibrium concept. That intrigue is accentuated by our desire to
  present the concept abstractly and independently. We do so because 
  we don't know in what other contexts the myopic 
 equilibrium concept can be applied.   
 \vskip.2cm
 \noi To bring into focus the  relationship between the 
 myopic and Nash equilibria, consider
 a three person game with simultaneous actions.
 %Represent the game so that Player One  chooses an action first followed by the other two players choosing their actions, but  without any  knowledge of what Player One  had done. 
 Let $I$ be the finite set of Player One's actions.  
One way to analyse this game is to define for  every distribution $p\in \Delta (I)$ %by the first player %on its actions  
a game $\Gamma (p)$ played by the
 second and third players based on their assumption that $p$ is the distribution by  which 
  the first player has acted.
 For every such $p\in \Delta (I)$, there will be a set of equilibrium strategies for the
  second and third players, and with them 
  corresponding payoffs for all three players in the game $\Gamma (p)$. If we return to the possible
  choices of the first player,
  represented by the set $\Delta (I)$,
  we recognise a correspondence of payoffs for Player One,
   determined
   by the $p\in \Delta (I)$ and the induced equilibria of the other two players. We can reformulate this
    as a one-player game with Player One as the only player. As a function of $p$, the corresponding
   payoffs for the first player will not   be affine;  in general  they will define  a correspondence.  We  could view this game as an
    optimisation problem -- the natural solution would be  that  the first player should 
    choose  the $p$ with the largest corresponding  payoff. With this approach, given a functional selection of payoffs  defined  on the $p\in \Delta (I)$,  one
    could see this optimisation
    as a kind of Nash equilibrium of a one player game.
     But this  optimisation  approach would in general  have no relationship 
     to the Nash equilibrium  of the original three player game! In a Nash equilibrium of a standard game defined by multi-linear functions, each action chosen with positive
     probability should share  a common  maximal  payoff among all the actions that can be taken. 
     But that in general %that 
 will  fail for the $p\in \Delta(I)$ that optimise the payoff for Player One in this one-player game;
      the different actions given positive probability could result in very different payoffs (and
     also  could  be dwarfed by the  payoff from an action given zero probability). 
      Rather the
     solution concept for the one-player game   directly relevant 
     to the Nash equilibria of the original three player game is that of the  myopic equilibria.

\vskip.2cm 
\noi  The rest of this paper is organised as follows. In the next section 
 we introduce the formal concept of myopic equilibria and prove its  existence 
 when the payoffs are continuous as  functions of the strategy spaces. 
 In the third  section we define a truncated game tree and prove 
 that all  composite games from truncated game trees 
  with certain structures have 
 equilibria. In the fourth  section we answer the question of A. Neyman 
 and speculate on closely related  applications. In the fifth and last section, we look at 
 examples and a possible future direction of research.

\section{Myopic equilibria} 

Nash equilibria are understood in terms of strategies that are 
 best replies to themselves. A best reply is a   strategy of a player  that can replace 
 that player's existing  strategy and maximise the payoff for that 
 player.
    Usually one 
 assumes that the set of strategies of a player is a compact and  convex set 
and that, given fixed strategies 
 of the other players,  
 the payoff to that  player is affine in its set of strategies. 
 If one assumes that the  payoff  
 function is concave in that player's strategies,
 the mathematics is similar, since  optimal responses (existent 
from the compactness of the strategy set)
  are realised on a convex subset. If the payoff function to 
 a player    
 is only  continuous with respect to his or her strategies, 
   one would not expect there to be  a Nash  equilibrium, which can  
 demonstrated with  simple  examples.  \vskip.2cm 
\noi  The idea that a strategy space is compact and 
convex  comes initially from 
 the assumption that    it 
  is the convex span of a finite set of actions. In this paper, we 
  keep  this assumption, though our definition of myopic equilibria could be
   generalised to a compact set of actions using  support sets. \vskip.2cm 
 
\noi{\bf Definition.} 
 Let $N$ be a finite set of players, and for each $n\in N$ let 
 $I_n$ be a finite set of  actions. Let $\Delta=\prod_{n\in N} \Delta (I_n)$ be  the strategy space for all the players. We say that ${x}\in \Delta$ is a {\em myopic equilibrium} \red{
 for a family of (payoff) functions  $\{w^n_i: \Delta \rightarrow \R\ |\  n\in N, i\in I_n\}$ if for all $n\in N$ and   $i\in I_n$ with $x ^n _i\ne 0$ one has  $w^n_i (x) = \max_{j\in I_n}  w^n_j (\red{x})$.
 
\skipv \noi
{\bf Convention.} Above and further, given $y\in \prod _{n\in N}\R ^{I_n}$ we denote by  $y^n$ the image of $y$ of under the natural projection onto $\R ^{I_n}$, and by $y^n_i$ the $i$--th coordinate of $y^n$, for $i\in I_n$. With the function $w: \Delta \to \prod _{n\in N}\R ^{I_n}$ satisfying $w^n_i(x)=(w(x))^n_i$, for all $x\in \Delta, n\in N$  and $i\in I_n$, we also say that $x$ is a myopic equilibrium ''for $w$'', instead of ''for $\{w^n_i\ |\  n\in N, i\in I_n\}$''.  }

\skipv 
\noi How does the myopic equilibrium concept compare with the 
 conventional  way to define a game and the conventional 
Nash  equilibrium concept?

\skipv\noi  With the myopic equilibrium concept there 
 are $|I_n|$ different payoffs for Player $n$, one for each of this player's actions, and they are functions o\red{n} the strategy space $\Delta$.  From these payoffs, one can define a functions $g^n$  from $\Delta$ to  $\red{\R}$ for each player $n$ in the canonical  way, by 
 $g^n (\red{x}) := \sum_{i\in I_n} \red{x}^n_{\red{i}}  w^ n_i(\red{x})$. Such functions  are not necessarily affine or concave in  the strategies  of a player. Starting with such functions $g^n$, 
 there will always be at least one way  to define corresponding
 functions $w^ n_i$ for
 the $i\in I_n$ that so  induce  the $g^n$ as above, namely  
 to define $w^ n_i(\red{x})$ to be $g^n(\red{x})$ for every 
 $i\in I_n$. By defining the $w^ n_i$ in this way  every
 point in $\Delta$ is a myopic 
 equilibrium, and that  is not interesting.  The interest in 
myopic equilibria lies entirely with 
  how the payoffs are defined for the individual actions.
 One must guarantee minimally that whenever 
 $\red{x}$ calls   for Player $n$ to  choose an  action 
 $j\in I_n$ with certainty it follows that    $w^ n_j (\red{x}) $ must 
 equal $ g^n (\red{x})$, but beyond that there will be many ways to 
 define the $w^ n_j$. 
     \skipv   
\noi  If the payoffs for all players are multilinear functions, %e.g. defined  through  multi-dimensional matrices, 
 one could say also that there are $|I_n|$ different 
 payoffs for each player $n$,
 defined   however   on the smaller set 
  $\red{\{i\} \times } \prod_{j\in N\backslash \{ n\}}
 \Delta (I_j)$ for each choice $\red{i\in I_n}$ of action in $I_n$.
  In this  special case,  %  when  the payoff to each player is defined in the conventional  multi-linear way, 
  a myopic equilibrium is the same as a Nash equilibrium. 
But when the payoffs to a player are not so defined, the two 
 equilibrium concepts can differ greatly, as we see 
 in examples \red{in \S 5}. 
 
 \skipv \noi There is a    concept of  {\em local} 
 equilibrium, a member ${x}$ 
 of $\Delta$ such that for every  $n$ the strategy ${x}^n$ 
 of player $n$ defines a local maximum of this player's payoff function. 
See Biasi and Monis (2013) for such an alternative 
  concept in the context of differentiable payoff functions. 
 However this concept of local equilibrium is still based 
 on functions $g^n$ defined on $\Delta$, without 
 necessarily separate functions $w^ n_i$ defined for each action, as
 described above.  We will see later from an example that local and myopic
  equilibria can be very different. 
\skipv 
  
We postpone  until later
discussing examples of myopic equilibria and pass  to establishing
some of their properties. We  show first that the myopic
equilibrium concept is  amenable  to a version the Structure Theorem of
Kohlberg and Mertens  (1986). 

\Theorem  Let ${\cal{W}}$ be a finite dimensional vector space of continuous functions defined on    $\Delta=\prod_{n\in N} \Delta (I_n)$ with values in \red{$\R ^I=\prod_{n\in N}\R ^{ I_n}$}{\red{.}} 
Assume that  ${\cal{W}}$ contains all the constant functions. 
Let $E$ be the \red{subspace} of ${\cal{W}} \times \Delta$  such that  $(w,\red{x})$ is in $E$ if and only if $\red{x}$ is a myopic equilibrium for $w$.  Then there exists a 
\red{homeomorphism $\phi$ of ${\cal{W}}$ onto $E$ whose post--composition with the projection to~${\cal{W}}$ is properly homotopic to the identity.}  

\Remark 
A homotopy $H: {\cal {W}}\times [ 0 ,1]\to {\cal W}$ being {\sl{proper}} means that $\inf_{t\in [0,1]}\|H(w,t)\|\to \infty$ as $\|w\|\to \infty$. A homeomorhism as asserted in Theorem 1 necessarily extends to an embedding of the one--point compactifications $\widetilde{W}$ of $W$ into that of ${\cal W}\times \Delta$, whose composition with the projection to~$\widetilde{W}$ is homotopic to the identity mapping of the sphere~$\widetilde{W}$. 
\skipv 

In the proof of this and the next theorem we'll use a property of a standard retraction of an euclidean space $\R ^J$ onto the probability simplex $\Delta (J)$.

\Lemma Let $J$ be a finite set. Then, there exists a continuous function $r_J: \R ^J\to \Delta (J)$ such that, given $x\in \Delta (J)$ and $y\in \R ^J$, condition $r_J(x+y)=x$ holds true if and only if $y_i =\max _{j\in J} y_j$ for all $i\in J$ satisfying $x_i\ne 0$. 

\skipv \noi {\bf Proof:} 
For each non-empty  $I\subset J$ we consider  $\Delta (I)$ as a face of  $\Delta (J)$ and define $Y_I=\{y\in \R^J: y_i=\max _{j\in J} y_j \ {\rm if} \ i\in I\}$. Observe that  the sets $Z_I=\Delta(I)+Y_I$ form  a closed  cover of $\R^J$. Since for any $z \in Z_I$ there are unique $x \in \Delta(I)$ and  $y \in Y_I$ such that $z=x+y$,  
we can define the projection  $\pi_I: Z_I \to \Delta(I)$  by $\pi_I(x+y)=x$, where  $x \in \Delta(I)$ and  $y \in Y_I$.
Note that for any two  non-empty subset $I$ and $I'$ of $J$,  $\pi_I$ and $\pi_{I'}$ coincide on $Z_I \cap Z_{I'}$. 
One can check that the map  $r_J: \R ^J\to \Delta (J)$ defined by the projections $\pi_I$ satisfies the assertion of the lemma.
~\hfill $\Box$  

\Remark It can be shown that $r_J$ is the nearest--point retraction with respect to the euclidean norm. (We don't use this here.)  

\skipv \noi {\bf Proof of Theorem 1:} 
Let $\r:=\prod _{n\in N}r_n : \R ^I\to \Delta$, where each $r_n\, (n\in N)$ is the corresponding mapping of $\R ^{I_n}$ onto $\Delta (I_n)$ given by Lemma 1 for $J=I_n$.
We divide the proof into 4 steps. 

{a)} As an immediate consequence of the definition of myopic equilibria it follows that a point $x\in \Delta$ is a myopic equilibrium for a function $w: \Delta\to \R ^I$ if and only if $\r   (w(x)+x)=x$, i.e., iff $\r ((w+x)(x))=x$. 

b) By a), $(w',x)\mapsto (w'-x, x)$ is a homeomorphism of $E':=\{(w',x)\ |\ w'\in {\cal{W}}\mbox{ and }r(w'(x))=x\}$ onto $E$, and as a map into ${\cal{W}}\times \Delta $ it is properly homotopic to the identity on $E'$ via the homotopy $\left((w, x),t\right)\mapsto (w-tx, x)$. It hence remains to construct a homeomorphism $\phi': {\cal{W}}\to E'$ satisfying the claim of the Theorem with $E$ and $\phi$ replaced by $E'$ and $\phi'$, respectively. 
\footnote{It is worth remarking that if it were the case that each $w\in {\cal{W}}$ was constant, as in Kohlberg and Mertens (1986), then one could finish this proof by letting $\phi '(w)=(w, r(w))$.} 
 
c) We now fix $x_0\in \Delta $ and define maps $\phi ': {\cal{W}}\to {\cal{W}}\times \Delta $ and $\psi ' : E' \to {\cal{W}}$ by the formulas (the composition signs are to be omitted):   
\begin{equation} \phi ' (w)=(w+w(x_0)-w\c   \r \c    w(x_0), \r \c     w(x_0))\, ,\end{equation} 
\begin{equation} \psi ' (w,x) =w-w(x_0)+w(x)\,.\end{equation} 
A direct verification shows that $\phi '({\cal{W}})\subseteq E'$ and $\psi ' \phi'$ and $\phi' \psi ' $ are identities on ${\cal{W}}$ and on $E'$, respectively.  Hence, $\phi'$ is a homomorphism of ${\cal W}$ onto $E'$.

d) The composition of $\phi '$ with the projection to ${\cal W}$ is given by the formula $w\mapsto w+w(x_0)-wrw (x_0)$, and we define a homotopy $H$ joining it to the identity by the formula $H(w,t)=w+t(w(x_0)-wrw (x_0))$. To show that $H$ is proper let us equip  ${\cal W}$ with the norm $\| w \| _{\sup}:=\sup_{x\in \Delta} \| w(x)\| $ induced by a norm $\| \, \| $ on $\R ^I$. Suppose, a contrario, that there exist $w_k\in {\cal W}$ and $t_k\in [0,1]\, (k\in \N)$ such that $\|w_k\|_{\sup} \to \infty$ and $\sup _k\| H(w_k, {t_k})\|_{\sup} <\infty$. On dividing the latter by $\|w_k\|_{\sup}$ and letting $u_k:=w_k/\|w _k\|_{\sup}$ we infer that 
$$ u_k+t_k\left(u_k(x_0)-u_krw_k(x_0)\right) \to 0\mbox{ as }k\to \infty.$$ 
By compactness of $\{w\ |\ \|w\|=1\}\times [0,1]\times \Delta$, the sequence  of triples %$\left((u_k, t_k, rw_k(x_0))\right)_{k=1}^\infty$ 
$(u_k, t_k, rw_k(x_0))$ has a cluster point, say $(u_0, t_0, y_0)$.
%we may take a sequence $(k(i))_{i=1}^\infty$ such that the sequences $(u_{k(i)}), (t_{k(i)})$ and $(rw_{k(i)}(x_0))$ converge respectively to $u_0, t_0$ and $y_0$, say. 
Hence we get
\[ u_0+t_0 \left (u_0(x_0)-u_0(y_0) \right )=0.\]
The second summand above being a constant function it follows that so is~$u_0$. Thus $u_0(x_0)-u_0(y_0)=0$ and next $u_0=0$. However, $u_0$ is a cluster point of the sequence of norm 1 vectors $u_k$, % $\|u_0\|_{\sup}=\lim _{i\to \infty}\|u_{k(i)}\|_{\sup}=1$ 
and  this contradiction establishes the properness of $H$ and completes the proof.~\hfill $\Box$   

\Remark The homeomorphism $\phi: W\to E$ constructed above has additionally  the property that for every $w\in {\cal{W}}$ the ${\cal{W}}$--component of $\phi (w)$ differs from $w$ by a constant function (i.e., a vector of $\R ^I$) whose norm is bounded by $2\| w\| _{\sup}+\delta$, where $\delta =\sup _{x\in \Delta}\| x\|.$
Also, $\|\phi ^{-1}(w,x)-w\| \le 2\| w\| _{\sup}+ \delta $ for $(w,x)\in E$. 

\skipv We also have a version of Nash's Equilibrium Existence Theorem. It is convenient to formulate it with an expanded definition of myopic equilibria in mind, when on $\Delta $ one has a multi--function $W$ (rather than a single--valued function $w$). 

\skipv \noi{\bf{Definition.}} Let to each $x\in \Delta$ be assigned a set $W(x)\subset \R ^I:= \prod_{n\in N}\R ^{I_n}$.

i) We say that ${x}\in \Delta$ is a {myopic} equilibrium for the multifunction $W:\Delta\to \R ^I$ if there exists a point $y\in W (x)$ such that whenever $n\in N$ and $i\in I_n$ satisfy $x^n_i\ne 0$, then $y_i =\max _{j\in I_n}y_j$. 

ii) If each set $W(x) $ is of a product form  $W(x)=\prod_{n\in N}\prod _{i\in I_n} W^n_i(x)$, where $W^n_i(x)\subset \R$, then in place of ''for the multifunction $W$ '' we also say above ''for the family of multifunctions $\left ( W^n_i\right ) _{n\in N, i\in  I_n}$ ''. 

\Theorem  Let $W$ be a multifunction on $\Delta$ which takes values in non-empty, closed,  convex subsets of $\R ^I$ and is upper--semicontinuous (meaning that $\{x\in \Delta \ |\ W(x)\cap  
K\ne \emptyset\}$ is closed in $\Delta $ whenever $K$ is closed in $\R ^I$). Then, there exists a myopic equilibrium for ${W}$. 

\skipv 
\noi {\bf Proof:}  If $W$ is single--valued and continuous, denoted now by $w$, then by Brouwer's Theorem the mapping $\Delta\ni x\mapsto r(w(x)+x)\in \Delta$ has a fixed point~$x_0$. (Here, $r$ is that from the proof of Theorem 1.)  By part a) of that proof, $x_0$ is an equilibrium for $w$.

In the general case we put a norm $\| \ \|$ on $\R ^I$. %By [....],  
For each positive integer $k$ there exists a single-valued continuous function  $w_k: \Delta \to \R ^I$ such that given $x \in \Delta$ we have $\|y- w_k(x')\|+\|x-x'\| < {1\over{k}}$ for some $x' \in \Delta$ and $y \in W(x)$. By the special case above, for each $k$ there exists a myopic equilibrium $x_k \in \Delta$ for the function $w_k$. Then, an accumulation point of the set $\{x_k\}_{k=1}^\infty \subset \Delta$ is a myopic equlibrium for  ${W}$. \hfill $\Box$

%\vskip2cm
\section{Game Trees and Incomplete Information} 

\noi We have to modify the concept of a finite game tree \red{(Kuhn (1953), cf. Hart (1985))} so that  the end points of the game are  states  for a continuation process, be
it a follow-up game or something else.  We call this  modification a {\em truncated game tree}. It involves 
 removing the final payoff from what  conventionally is  defined 
 to be a game tree. The term is justified because any shorter truncation of a truncated game tree is also a truncated game tree.  With our  application, instead of a payoff determined 
 by the end point  there is  a continuation payoff   determined by 
 the induced conditional probability
 distribution on the end points known in common, (which could be interpreted as
    a kind of  subgame). 
 But these continuation payoffs and their relationship 
 to the conditional probabilities are exogenous   to 
 the truncated game tree.
   \skipv 
   
\noi The main inspiration is any game for which all  players observe 
all  actions taken, however they don't observe the decision process behind 
 those actions. The distinction can be strong with games of incomplete 
 information, where a player can posses a secret   and makes its 
 behaviour dependent on that secret.   As with  
 poker, though one observes completely the behaviour of 
 other players, it is the relationship between  their private knowledge 
 and their behaviour that one needs to understand as a player.    
 \vskip.2cm
 \noi 
A game tree has vertices $V$ and directed edges or arrows between the vertices.
 Its  vertices $V$ can be broken down into two  types,  
    nodes and end points.    $E$ is the set of end points 
     and every path of arrows
    starts at the root and ends at an end point,
 with  each end point determining
    a unique such path of arrows. 
  The set $D$ of nodes  is   the  subset
  $ V\backslash E$ and these are the vertices
  (except for the root $r$) to which comes exactly one arrow
    and from which, without loss of generality,  come at least two 
    distinct  arrows. \vskip.2cm 

  %  \red{ Bob, do we need or use this additional assumption?}
 %\blue{I think the more detail the better, since we want to go beyond
  %  game theorists, who are the only ones who would be familiar.} 
%\skipv 

    \noi For each player $n\in N$  there is a subset $D_n\subseteq D$ such that 
 $\forall i \not= n\ D_i \cap D_n = \emptyset$. Define $D_0$ to be  
  the set $D\backslash ( \cup_{n\in N} D_n) $. To every player $n\in
  N$ there is a partition ${\cal P}_n$ of the set $D_n$.
   \skipv
 
\noi For  every 
 $W\in {\cal P}_n$ with $W\subseteq D$  there is a corresponding set of actions  $A_W^n$ such that there is a bijective relationship between $A^n_W$ and the  arrows leaving \black{every $v\in W$.}
 For every $v\in D_0$ there is a probability distribution 
 $p_v$ on the arrows leaving the node   $v$, and therefore also 
 on the nodes following directly after $v$ in the tree. 
\skipv

%\blue{I think we should stick with $E$ the end points, and just not be explicit% about the edges of the directed graph. The
 %  graph structure is used only to describe it initially, isn't really used in %any proofs.} 
\noi  At any node $v\in W \in {\cal
  P}_n$ only the player $n$ is making any decision, and this decision
  determines completely which vertex follows $v$.  At the nodes $v$  
 in $\red{D }_0$ nature is making a decision, according to $p_v$,  concerning
  which vertex follows $v$. If the game is at the node $v\in D_n$ and
  $v\in W \in {\cal P}_n$ then Player $n$ is informed   that the node is
  in the set $W$ and that player has no additional information, so that 
 inside $W$ player $n$ cannot distinguish between nodes within $W$.

\skipv 
\noi Notice that any simultaneous move game can be so modeled, by 
 choosing any order of players and giving all players 
 indiscreet partitions. \skipv 

 \noi With conventional game trees, we assume that once the set $E$ of
 end points is reached that the game is over and the players learn
 the outcomes. But  a truncated game tree may be a prelude to
 further activity, or
 the payoffs may be  exogenous to the truncated game tree.
 We may need to define the knowledge of the players
  at the set $E$. For each player $n\in N$ 
 let ${\cal Q}_n$ be a partition on  $E$.  
 Let  ${\cal Q}:= \wedge_{n\in N} {\cal Q}_n $ 
be the join partition on $E$, meaning the 
  finest  partition such that for every $n\in N$ every member of
 $ {\cal Q}_n$  is contained in 
 some member of ${\cal Q}$. The partition ${\cal Q}$ corresponds 
 to the concept of common knowledge, meaning that  a member $C\in {\cal Q}$ 
 is what the players know in common whenever $e\in C$ is the 
  resulting end point. If there is a continuation game, the corresponding 
 set $C\in {\cal Q}$ defines the appropriate subgame.
\skipv

%\blue{right after the partitions on $E$ are explained we should put all the discussion of perfect recall. I know you want
 % it all the way at the end, because we don't really use it till the last section. But I disagree. The clever reader knows
%already, without being a game theorist, that there is something wrong with a player whose partition at an earlier stage
 % implies greater understanding than that implied at a later stage.  So as soon as the last partitions are defined,
%namely the ${\cal Q}_n$, one should discuss perfect recall.} 

\noi {\bf Definition:} The truncated game  tree   has {\em perfect recall} 
 for a player $n$   if all paths leading to a  partition member  %$W$ 
in either ${\cal P}_n$ or $ {\cal Q}_n$ 
  pass through the same previous partition  sets in 
${\cal P}_n$ in   the same order and without repetition.  \vskip.2cm 
\noi    
 Though much   is stated and proven without the assumption of perfect recall, it would 
 be difficult to understand  the  relevance of most  of  what follows without the assumption of perfect recall.  
\vskip.2cm

\noi 
 For every player $n\in N$ let  $S_n$ be the finite set of pure decision 
 strategies of the players   in the truncated  game tree, 
by which we mean a function that decides, at every set $W$  in ${\cal P}_n$, 
%such that $W$ is also  contained in  $D_n$,  
deterministically which 
  member of $A^n_W$ should be chosen.  
If each such  $A^n _W$ has cardinality 
 $l$ and there are $k$ such sets then the cardinality of 
 $S_n$ is $l^k$.    \skipv

\noi Now we define a new payoff structure  
    from the truncated game tree and continuation  
 payoffs.   
 For any $C\in {\cal Q}$  let there be a correspondence 
 $F_C \subseteq \Delta (C) \times {\R}^{C\times N}$  of {\em continuation} payoffs 
%As it is finite,  we perceive $E$ as a compact space 
and  for every $n\in N$ and $e\in E$  let  $g^{e,n}: \red{\R} \rightarrow \red{\R} $ be \red{a function}. \red{\footnote{The  application in \red{\S 4}  will be $g^{e,n}(t)= \lambda r^n_e + (1-\lambda) t$ for  some $0 < \lambda < 1$ where   $r_e\in \red{\R}^N$ is  a payoff vector  associated  with the end point $e$.
 If the multifunctions $F_C$ were constant,  that is $F_C (p) = F_C (p')$ for $p, p'\in \Delta (C)$, the payoff structure we define would not 
 be different from that of a conventional game tree.} }
\vskip.2cm

\noi 
For every $x=(x^n)_{n \in N}  \in \Delta := \prod_{n \in N} \Delta(S_n)$, 
by $p_x$ we denote the probability distribution on $E$ defined by $x$, 
and for  $C\in {\cal Q}$  with  $p_x (C) > 0$  by $P_x (\, \cdot \,| C)$  the  conditional probability on $C$ induced $x$.  
For $s \in S_n$, by $x^s$ we denote the element of $\Delta $ obtained from $x$ by replacing $x^n$ by $s$.

We say that a vector $(y^n_s)_{n \in N, s\in S_n}$  is  {\it proper for} %with respect to} 
$x \in \Delta $ if  
$$ y^n_s = \sum_{e \in E} p_{x^s}(e) g^{e,n}(\nu^{e,n}),  \   n \in N,    s\in S_n \  , $$
for some  $\nu = (\nu^{e,n})_{ e \in E, n\in N}  \in \R^{E \times N}$  such that   
 for each   $C \in {\cal Q}$, the image of $\nu$ under the natural projection to $\R ^{C\times N}$ belongs to 
 $F_C (P_x(\, \cdot \, | C ))$ if  $p_x(C) > 0$ or else, if $p_x(C)=0$,
  belongs to some  
  $F_C (q)$ for some $q\in \Delta (C)$ as determined in some way 
   by $x$.\vskip.2cm 

   \noi The term  ''proper values'' refers to the fact
   that the continuation  payoff corresponds to  the conditional 
 probability distribution,
 given that it  is  well defined.  When the  conditional probability is not well defined, meaning that a set $C\in {\cal Q}$ has reached that shouldn't have been 
 reached according to ${x}$,  the continuation payoff corresponds to some distribution on $C$.    That zero probability of reaching $C$ according to ${x}$  implies that somebody has acted in 
 an inappropriate way and the use of such a continuation payoff could be interpreted as punishment. However there are problems 
with seeing such a continuation payoff as the punishment of some  particular player, and this is discussed below.   %\skipv 
 
  %$\nu_C := (\nu^{e,i})_{ e \in C, i\in N} \in  F_C (\Delta(C))$ for each   $C \in {\cal Q}$ and  $\nu_C \in  F_C (P_x(\, \cdot \, | C )$ if $p_x(C) > 0$. 

%\bigskip 

\Theorem Let $F_C : \Delta(C) \to \R^{C \times N}$,  $C\in {\cal Q}$,  be upper semi-continuous correspondences with non-empty convex values,   and  $g^{e,n}: \R \to \R$,  $(e,n) \in E \times N$, continuous increasing functions. 
Then there exist $x \in \Delta$ and a vector $(y^n_s)_{n \in N, s\in S_n}$  proper for  $x$ and such that 
$y^n_s \ge y^n_t$ for all  $n \in N$ and all $s,t \in S_n$ with $x^n_s > 0$. 
\vskip.2cm

\noi {\bf Proof:} Let  $\epsilon>0$ be given  and let $B$ be a positive quantity larger than any payoff from 
 the correspondences $F_C$. 
 For each   $C\in {\cal Q}$   there is a function  $\phi_{C,\epsilon}   : \Delta (C)
 \rightarrow {\bf R}^{C\times N}$ that is a  continuous $\epsilon$ approximation of  $F_C$.   
 If $p_{x }(C) \ge \epsilon$  then define  $\lambda_{x,\epsilon} (C) =1$,    
   and  define $\lambda_{x,\epsilon} (C)=  \frac {p_{x} (C)} {\epsilon}$   if  $ p_{x} (C) \le \epsilon$.  

 For every $x\in \Delta$, $n \in N$  and $e \in E$ let  
$$
\tilde f^{e,n}_\epsilon (x)= 
g^{e,n}\big( \lambda_{x,\epsilon} (C) \phi^{e,n}_{C,\epsilon}(P_{x} (\, \cdot \,| C)) +  (1-\lambda_{x,\epsilon} (C) ) 2B\big) \, ,
$$
where $C$ contains $e$ and  $\phi^{e,n}_{C,\epsilon}$ is the $(e,n)$-th coordinate of $\phi_{C,\epsilon}$. 
If $p_x(C)=0$ then $\tilde f^{e,n}_\epsilon (x)=g^{e,n}(2B)$. 
For each  $s\in S_n$,  we let 
$$ 
\tilde f^n_{s,\epsilon} (x)= \sum_{e \in E} p_{x^s}(e)\tilde f^{e,n}_{\epsilon}(x) \,  . 
$$
 Notice that   the $\tilde f^n_{s,\epsilon} (x)$  are continuous in $x$.
By Theorem 2,  for each $\epsilon > 0$ there exist a myopic equilibrium, say $x(\epsilon)$, 
for the family $(\tilde f^n_{s,\epsilon})_{n\in N, s\in S_n}$. 
Thus 
$$
\tilde f^n_{s,\epsilon} (x(\epsilon)) \ge \tilde f^n_{t,\epsilon} (x(\epsilon)) \  \ 
{\rm for} \ s,t \in S_n \ {\rm with} \ x(\epsilon)^n_s >0 \, .
$$ 

Observe that for some sequence $(\epsilon_k)_{k \in \N}$ converging to 0 we have: 

\medskip

(a) the sequence $\big(x(\epsilon_k)\big)_{k \in \N}$ converges to some $\tilde x \in  \Delta$,

\medskip

(b) for each $n \in N$ and $s \in S_n$, the sequence $\big(\tilde f^n_{s,\epsilon_k} (x(\epsilon_k)) \big)_{k \in \N}$   converges to some ${\tilde y}^n_s$,

\medskip

(c)  for each $e \in E$ and $n \in N$, the sequence  $\big({\tilde \nu} ^{e,n}_{\epsilon_k}  \big)_{k \in \N}$, where 
$$ 
{\widetilde \nu} ^{e,n}_{\epsilon_k} =  
 \lambda_{x(\epsilon_k),\epsilon_k} (C) \phi^{e,n}_{C,\epsilon_k}(P_{x(\epsilon_k)} (\, \cdot \,| C)) + 
 (1-\lambda_{x(\epsilon_k),\epsilon_k} (C) ) 2B \, , 
$$
converges to some ${\widetilde \nu} ^{e,n}$. 

Note that for all $n \in N$ and $s \in S_n$ we have
$$
\tilde y^n_s = \sum_{e \in E} p_{{\tilde x}^s}(e) g^{e,n}(\widetilde \nu^{e,n})  \, .  
$$
Observe that if ${\tilde x}^n_s > 0$, where $s \in S_n$, then $x(\epsilon_k)^n_s > 0$ for almost all $k$. 
It follows that
$$
\tilde y^n_s  \ge \tilde y^n_t  \  \  
{\rm for} \ s,t \in S_n \ {\rm with} \ \tilde x^n_s >0 \, .
$$ 
Note also that the sequence $\big(p_{x(\epsilon_k)}(C)\big)_{k \in \N}$   converges to $p_{\tilde x}(C)$. 

Suppose that $p_{\tilde x}(C) > 0$.  Then $p_{x(\epsilon_k)}(C) > \epsilon_k$  for all sufficiently large $k$'s. 
For such $k$'s we have ${\widetilde \nu} ^{e,n}_{\epsilon_k} =  \phi^{e,n}_{C,\epsilon_k}(P_{x(\epsilon_k)} (\, \cdot \,| C))$  
for all $(e,n) \in C \times N$. 
Thus by (c), the sequence $\big({\widetilde \nu}_{C,\epsilon_k} \big)_{k \in \N}$,  where  
 ${\widetilde \nu}_{C,\epsilon_k} = \big({\widetilde \nu} ^{e,n}_{\epsilon_k} \big)_{e \in C, n \in \N}$,
converges to  ${\widetilde \nu}_{C}\in F_C(P_{\tilde x}(\, \cdot \,| C))$. 

Now, suppose $p_{\tilde x}(C) = 0$. 
If for almost all $k$,  $p_{x(\epsilon_k)}(C)=0$,  we choose an arbitrary  ${\nu}_C$ in  $F_C(\Delta(C))$.
Otherwise, as ${\nu}_{C}$ we choose any cluster point of the set 
$\{ \phi_{C,\epsilon_k}(P_{x(\epsilon_k)}(\, \cdot \, | C) \, | \, p_{x(\epsilon_k)}(C) > 0 \}$, 
which is also in $F_C(\Delta(C))$. 

Now, let $\nu = (\nu^{e,n})_{e\in E, n \in  N} \in \R^{E \times N}$ be such that the projection of $\nu$ onto $\R^{C \times N}$
 is the vector ${\widetilde \nu}_{C}$  if $p_{\tilde x}(C) > 0$ and the defined above vector  
${\nu}_{C}$  if $p_{\tilde x}(C) = 0$. 
For $n\in N$ and $s \in S_n$, we define 
$$
y^n_s = \sum_{e \in E} p_{{\tilde x}^s}(e) g^{e,n}(\nu^{e,n}),  \, .  
$$
Note that by the definition the vector $(y^n_s)_{n \in N, s\in S_n}$  is  proper for  ${\tilde x}$. 

Since  $\nu^{e,n} \le \tilde \nu^{e,n}$ for all $e \in E$ and $n \in N$, it follows that $y^n_s \le \tilde y^n_s$ for all $n\in N$ 
and $s \in S_n$. 
Now, for any $x\in \Delta$, if   $x^n_s>0$ for some  $s \in S_n$ and $p_x(C) = 0$ then  $p_{x^s}(C) = 0$.
It follows that if $\tilde x^n_s>0$ for some  $s \in S_n$ then $y^n_s = \tilde y^n_s$.
 Consequently, for $ s, t \in S_n$ with  $\tilde x^n_s>0$, we have $y^n_s=\tilde y^n_s \ge \tilde y^n_t \ge y^n_t$,
which completes the proof.  \hfill  $\Box$ 
\vskip.2cm
\noi  {\bf Remarks:}
a) The proof of the above theorem  has a resemblance to ``trembling hand'' arguments in Selten (1975), 
 however the mechanism for giving small probabilities to potentially 
 undesirable actions is very different.  \skipv

b) What the players observe 
 in common  is   some set $C$ in ${\cal Q}$.  Given that 
 they know 
  each other's strategies, the choice of $\red{x}$ in $\Delta$, 
  they know in common a conditional  probability 
 distribution on elements contained 
 in  the  set $C$ in ${\cal Q}$.  This doesn't mean that 
 each player knows only this about the payoffs,
 either his or her payoff or those 
 of others.  A player may learn much more, including potentially   
 exactly which $e\in C$ will  be reached for any given $C\in {\cal Q}$.  In such an   event the player 
 evaluates  his or her actions according to that exact knowledge of the end point $e$, however 
 knowing also that the payoff at  $e$ is determined by the induced common knowledge 
  distribution on $C$.    There is a similarity with poker, in which a player may know 
 that he or she has the winning hand, but that player's  betting strategy reflects an understanding 
 of what all players believe.  

\skipv 
c) It would be tempting to define the continuation payoffs  from the 
  $F_C$  always as those  from  a  game, 
 that is payoffs  generated by strategies. However we would then  
 require for all $e\in C\in {\cal Q}$ some determination 
 of a payoff for each player $n\in N$, including the case of 
some $W\in {\cal Q}_n$  given zero 
 probability by  the 
 relevant strategy ${x}$. There is 
 a problem with defining a player whose presence in the game has zero 
 probability and yet receives a payoff that could potentially torpedo  
 the equilibrium property. On the other hand, we did  
  need  to define such 
 payoffs, as we had to consider the payoff 
 consequences of decision functions chosen with 
 zero probability according to $\red{x}$ and make sure that 
 they did not profit the player in question over those decision 
 functions given positive probability.  \skipv 

d)   Also   tempting would be to interpret  
 the  landing at  a $C\in {\cal Q}$ that is given zero probability by
 the ${x}\in \Delta$   as the trigger of some  
 punishment of a player.
 With two-player games, if  only one player had deviated,  indeed  
   that  player can be held  responsible for bringing the play to the set $C$.  
 But with three  or more players, it may  be impossible to obtain 
 common knowledge of  
   which player 
 had brought the game to this forbidden subset.  Imagine the following example;  there are 
 three players $i=1,2,3$ and  each 
 player has three strategies, left, right, and centre, and each player is 
 required to play only centre. If all three
 players choose centre, then all three 
 players are informed of this fact. If Player $i$ chooses left then 
 Player $i-1$ (modulo 3) 
  is informed of this fact and if Player $i$ chooses 
 right then Player $i+1$ is informed   of this fact; and in 
 either case if Player 
 $i$ was the only disobedient player, the only information 
 that the  third other player receives is 
  that not all three players had chosen centre. Lets assume that Player $1$ 
discovers that one of the other  players was disobedient, but not 
 which one. There are two 
 possibilities, either Player $2$ played right or Player $3$ played left. 
 Players $2$ and $3$ could both maintain 
 that they were not disobedient.  The effective punishment 
 of Player $3$ may be very beneficial to Player $2$ 
 which could place an otherwise sound equilibrium in doubt, as then Player $2$ could deviate and 
 then claim that it was Player $3$ who deviated. With two players, this  problem 
 doesn't appear, because the two could punish each other. 
With the above theorem, there 
 is an implicit punishment through the choice 
 of some continuation payoff for all the players,
 but no explicit punishment strategies, which may  prove  problematic. \skipv

e) We  could have stated the theorem so that the 
multifunction of payoffs applies 
 only   to     all distributions that can be generated by strategies, but it would have made no difference. 
This is because 
    the set of distributions generated by strategies is closed, and  
    an upper-semi-continuous  multifunction with values as described  and defined on a closed subset of distributions can be extended to an analogous multifunction  defined  on all distributions.
    %(We use the fact that continuous functions  can be so extended, and so  the same holds for u.s.c. multifunctions that take values as above.)}

%\vskip2cm
\section{Games of Incomplete Information on One Side}

We return to Neyman's question.  
    There is a finite set $K$ of states of 
   nature. 
  Nature chooses a  
   state $k\in K$
 according to a commonly known probability on $K$, 
    and  
\noi  Player One, but not Player Two, 
   is informed of nature's choice.  
\noi  The finite  sets of moves for the players 
 are the same for all states, the set $I$ for Player One 
 and the set $J$ for Player Two. 
   After each stage of play, both players 
     are informed of each others' moves.
 The play is repeated indefinitely, and the chosen state remains constant  
  throughout play. 
\vskip.2cm

\noi 
 For every state $k\in K$ let $A^k$ and 
 $B^k$ be the payoff matrices of 
the two players   with $I$ indexing the rows 
  and $J$ indexing the columns. The   entries $a^k_{i,j}$ 
  and $b^k_{i,j}$ in $A^k$ and $B^k$ are
   the payoffs to the first and second players respectively. 
\noi     given that  the state is $k$,  
\noi  the move of Player One is $i$,  
\noi  and 
     the move of Player Two is $j$.
\vskip.2cm 

     \noi  The  strategies of the game are the same as those 
  described in Simon, Spie{\.z}, and Toru{\'n}czyk (1995) and 
 Aumann and Maschler (1995), though 
the payoffs are defined differently. 
  For the sake of completeness, we describe the strategy and payoff 
  structures below. 
 \vskip.2cm 
\noi A behaviour strategy of Player One  is  
  an infinite 
           sequence $\alpha=( \alpha^1, \alpha^2, \dots )$ 
       such that for each $l\ $ 
           $\alpha^l$ is a mapping from  
           $K\times (I\times J)^{l-1} $ to $\Delta (I).$    
  \vskip.2cm  
\noi            A behaviour strategy of Player 
           Two is 
  an infinite 
           sequence $\beta =( \beta^1, \beta^2, \dots )$     
       such that for each $l$ 
          $\beta^l$ is a mapping from  
           $(I\times J)^{l-1} $ to $\Delta (I).$  
     \vskip.2cm 
\noi        Let ${\cal I}$ and ${\cal J}$ be the set of behaviour strategies 
             of  Players One and Two, respectively. 
                       Define   
           the set of finite play-histories of length $l$ to be  
           ${\cal H}_l:   
           =K\times (I\times J)^l$, 
  and define ${\cal H}_l^k$ to be 
            the subset
\noi $\{ k\} \times (I\times J)^l$. 
\vskip.2cm 

\noi 
\noi  For any fixed 
$k\in K$,  
every pair of behaviour strategies   $\alpha\in {\cal I}$ and 
 $\beta \in {\cal J} $
 induces  a probability measure $\mu ^{l,k}_{\alpha, \beta}$ 
on     ${\cal H}_l^k$,   and with 
 the initial probability              
 $p_0$
  such a pair  
induces a probability measure $\mu ^l_{\alpha, \beta}$ 
on ${\cal H}_l$.
\vskip.2cm 
\noi  To define the payoffs, for both players $i=1,2$ there is a finite
 sequence $\lambda^i_1, \lambda^i_2,
 \dots , \lambda^i_n$ of non-negative real numbers such that
 $\lambda^i= \lambda^i_1 + \dots + \lambda_n^i$ and
 $0\leq  \lambda^i \leq 1$. For every $h\in {\cal H}_n$ with $h= (k,i_1, j_1, \dots ,
  i_n , j_n)$ define 
 $ f^1_n(h) $ to be $\sum_{l=1}^n \lambda^1_l a^k_{i_l,j_l}$ and 
  $ f^2_n(h)$ to be  $\sum_{l=1}^n \lambda^2_l b^k_{i_l,j_l}$. For
  every $h\in {\cal H}_m$ with
  $h=(k,i_1, j_1, \dots , i_m, j_m)$ define
  $\tilde f^1_m (h)$ to be $\frac 1m  \sum_{l=1}^m  a^k_{i_l,j_l}$ and
   $\tilde f^2_m (h)$ to be $\frac 1m  \sum_{l=1}^m  b^k_{i_l,j_l}$. 
\vskip.2cm 
\noi         An equilibrium is a pair of behaviour 
       strategies $\alpha \in {\cal I}$ and $ \beta\in {\cal J}$ such that 
       for every $k\in K$
\vskip.5cm 

 $$a^k=  \int _{{\cal H}^k_n} 
        f^1_n (h) d \mu^{n,k} _{\alpha , \beta } + (1-\lambda^1) 
   \lim_{m\rightarrow \infty}  
      \int _{{\cal H}^k_m} 
      \tilde f^1_m (h) d \mu^{m,k} _{\alpha , \beta }$$     
       and 
      $$ b^k=  \int _{{\cal H}^k_n} 
       f^2_n (h) d \mu^{n,k} _{\alpha , \beta } +
 (1-\lambda^2) \lim_{m\rightarrow \infty}  
       \int _{{\cal H}^k_m} \tilde f^2_m (h) 
d \mu^{m,k} _{\alpha , \beta }$$
  exist and 
     for every 
 pair $\alpha ^*\in {\cal I}$ and $\beta^*\in {\cal J}$                       
$$ \int _{{\cal H}_n} 
       f^1_n (h) d \mu^{n} _{\alpha^* , \beta } + 
(1-\lambda^1)\lim_{m\rightarrow \infty} \mbox { sup} \int _{{\cal H}_m} 
     \tilde f^1_m (h) d \mu^m_{\alpha ^* ,
      \beta }  \leq \sum_k p_0^k a^k \mbox { and } $$ 
$$ \int _{{\cal H}_n} 
      f^2_n (h) d \mu^{n} _{\alpha , \beta } +
 (1-\lambda^2)\lim_{m\rightarrow \infty} \mbox { sup} \int _{{\cal H}_m} 
       \tilde f^2_m (h) d \mu ^m _{\alpha , \beta ^*} \leq \sum_k p_0^k b^k.$$
\vskip.2cm 

\noi Such games as described above we call {\em Neyman } games, to 
 distinguish them from the conventional infinitely repeated games of incomplete information on one side,  introduced 
 in Aumann and Maschler (1995). If $\lambda^i=0$ for both $i=1,2$ then the game is the %conventional 
 one described there %introduced by Aumann and Maschler 
 and the above is the definition
 of an equilibrium of such a game. \vskip.2cm 
 \noi Notice the asymmetry in the behaviour strategies used to define equilibria. Player One's strategy uses knowledge of the state of nature, so the maximisation, relative to a fixed strategy
 of Player Two, can be performed on each state independently. Player Two's knowledge of the state of nature comes only from a  calculations of Bayesian
 conditional
 probabilities according Player One's chosen strategy and  the actions taken. \vskip.2cm
 \noi 
With regard to the infinitely repeated game %introduced 
in   Aumann and Maschler (1995),  these authors with the help of R. Stearns  introduced a
 solution concept known as a {\em joint plan}. For any 
 $p\in \Delta (K)$ define $a^* (p)$ to be the value 
 of the zero-sum game defined by the matrix 
 $A(p):= \sum_{k\in K} p^k A^k$, where  $p^k$ is the probability that $p$ gives 
 to the   state $k\in K$. Likewise define $b^*(p)$ to be 
 the value of the zero-sum  defined by the matrix 
 $B(p) := \sum _{k\in K} p^k B^k$. 
 A   vector $x\in {\bf R}^K$ is  {\em individually rational}  for
       Player One if 
  $x\cdot q\geq a^*(q)$ for 
             all $q\in \Delta (K)$. A pair $(r,p)\in {\bf R}
 \times \Delta (K)$ is individually 
 rational for Player Two if $r\geq \mbox {vex} (b^*) (p)$, where 
 $\mbox {vex} (b^*)$ is the unique  convex function satisfying 
 $\mbox {vex} (b^*) \leq b^*$ and $\mbox {vex}(b^*) \geq f$ for all 
 convex functions $f$ such that  $f\leq b^*$. 
For   every $\gamma\in \Delta (I\times J)$ 
     define    $\gamma A\in {\bf R}^K$ by                
       $$(\gamma A)^k := 
      \sum_{(i,j)\in I\times J}\gamma^{(i,j)} A^k(i,j)$$  
     and define $\gamma B$ likewise. 
                A  joint plan  
  for an initial probability $p_0$ 
        is \vskip.2cm  
     
\noi          (1) a finite subset of 
         probabilities $V\subseteq \Delta (K)$ such that the  convex hull  
          of $V$ contains 
          the initial probability $p_0$, \vskip.2cm 
\noi  
(2) for every $v\in V$ a $\gamma_v \in \Delta (I\times J)$,\vskip.2cm  
\noi  
(3) for some finite $n$ 
a finite set $T\subset I^n$ of signals   
 in bijective relation to the set $V$ 
 and a state dependent choice of an $s\in T$  
   performed by Player One
 such that the signal $s\in T$ implies by Bayes rule a 
 conditional probability on the set $K$ equal to its corresponding
         member in $V$. 
\vskip.2cm 
\noi  
(4) if the signal $s$ chosen corresponds to 
 $v\in V$, 
an agreement between the players to play through the rest of the game a 
 deterministic sequence of pairs of actions $((i_1, j_1), (i_2,j_2), \dots)$ 
 such that in the limit the distribution $\gamma_v$ is obtained, and
         \vskip.2cm 
\noi  
(5)  punishment strategies of the two players to be implemented in the 
 event that a player does not adhere to the agreed upon sequence of 
 actions.       
\vskip.2cm

\noi Aumann and Maschler showed that 
a joint plan describes an equilibrium of the undiscounted game  
if   
there is an individually rational $y\in {\bf R}^K$ such that  
 for every $v\in V$ the following holds:  \vskip.2cm 
\noi (1) $(\gamma_v B)\cdot v \geq \mbox {vex} (b^*)(v)$, \vskip.2cm  
\noi (2) $\forall k\in K \ (\gamma_v A) ^k= y^k$ if $v^k > 0$,
\vskip.2cm  
\noi (3) $\forall k\in K \ (\gamma_v A)^k \leq y^k$ if $v^k=0$.   
  \vskip.2cm

  \noi If necessary, Player One  is 
punished according to a strategy of Player Two such that simultaneously 
  for every $k\in K$ 
 Player One is held down to no more than  $y^k$. This ability of Player  Two 
 is based on a theorem of D. Blackwell (1956). \vskip.2cm
 \noi The punishment of Player Two centers on the conditional probability of the states of nature as implied by
 the actions taken and the chosen strategy of Player One. There is a qualitative difference between
 the punishment of the two players. The punishment of Player One is absolute with a quantity determined for
 each state simultaneously. The punishment of Player Two is relative to a conditional  probability distribution on the states of nature.
 The need to calculate  payoffs according to expectation    gives the effective punishment. \vskip.2cm  
%%%%%%%%%%%%
 
 \noi 
 The equilibrium payoffs of a joint plan equilibrium
 is the pair $(x,y)\in {\bf R}^K \times {\bf R}^K$ such that
 for every $k\in K$ the value $x^k$ is what the first player
 gets in average expectation in the limit at
 the state $k$ and $y^k$ is what the second
 player gets in average  expectation in the limit
  at the state $k$.  Notice from the structure
  of a joint plan that these values are well defined. 
  \vskip.2cm
  \noi 
 %I looked at the paper of Hart again. The statement of the main theorem is on page 125, and the mechanism used is called the jointly controlled lottery. The main theorem on 125 goes far beyond what we would use, and that is why it is difficult to pin point where Hart proves what we need. The direction of the proof that we need starts on page 135, but that doesn't help that much. More direct help comes from the Aumann Maschler book Repeated Games with Incomplete Information. On page 272 they define the jointly controlled lottery and later, on page 276, they state their theorem 7.2 which is explicitly what we need. As you know, the book is prior, as it comes from 1968 and not really its publication date. 
  
Hart (1985)
  showed that if  $(x_1,y_1), (x_2, y_2)\in {\bf R}^K \times {\bf R}^k$ are
  both equilibrium payoffs of two distinct joint plan equlibrium corresponding to the same initial
   probability distribution on the states, then 
  for every $0\leq \lambda \leq 1$
   there is an equilibrium of the game that delivers
  expected payoffs of
  $\lambda (x_1, y_1) + (1-\lambda) (x_2, y_2)$. The players accomplish this
  through a {\em jointly controlled lottery}, a way for the players to choose
  one or the other joint plan equilibrium through an initial phase of
  independent random behaviour. See also Aumann and Maschler (1995) for
   an explanation of a jointly controlled  lottery.

\noi Now we apply Theorem 3 to prove the following theorem. \vskip.2cm 
 
\Theorem The above question 
 of A. Neyman is answered in the affirmative,
 meaning that every  Neyman game  has an  equilibrium.    \vskip.2cm 
 \noi {\bf Proof:} We have to define the truncated game tree,
  the mixed strategy space $\Delta$, 
 the partitions ${\cal Q}_i$
 on the end points of this tree, the continuation vectors
 $F_C$ for every $C\in {\cal Q}_1 \wedge {\cal Q}_2= {\cal Q}$,
 the payoff functions $g^{e,i}$ for the players $i=1,2$, and also what continuation payoff is chosen when a $(\sigma, \tau)\in \Delta$ means that
 the corresponding  $C$ will be reached with zero probability.
 
  \vskip.2cm 
\noi The first $n$ stages of a Neyman game define the  truncated 
game tree $\Gamma_n$ for
which $E:= K \times (I\times J)^n$ are the end points. The truncated
 game tree has  
  $2n+1$ levels of play, the first level  being Nature's choice and 
 the $2n$ following levels being alternations between
 Player One's and Player Two's choices of actions. 
   The first to move 
 is Nature, choosing some $k\in K$.   After Nature's choice, Player One 
 has a partition consisting of $|K|$ different singletons, representing 
 a complete knowledge of Nature's choice. This is followed by 
 an action of Player Two, for which Player Two 
 has only one partition member for this stage of play, meaning 
 that Player Two has no information on which to base his  choice 
 of action.  For every $m< n$, at the  
 conclusion of the $m$th stage
 (meaning that $2m+1$ actions have been performed,
 $m$ by both  Players One and Two and the first by Nature)
   Player One's partition consist of  
 the singletons of $K\times (I\times J)^m$, which are used 
 to determine 
 Player One's $m+1$st action, followed by 
  partition elements for Player Two (to determine  his $m+1$st action)
   defined  by 
 the different members of $(I\times J)^m$ 
(meaning that Player Two saw the first 
 $m$ actions of Player One but not the $m+1$st action).
 The partition ${\cal Q}_1$ on $E$ for Player One consists of the  
 $|K| \cdot |I|^n \cdot |J|^n$ many  singletons (meaning that
 at the conclusion of the truncated game tree Player Two does learn what
  Player One did in the last stage of that tree).
  The partition ${\cal Q}_2$ on $E$ for the second 
 player consists of the sets of size $|K|$ of the form $K \times \{x\}$ for 
 all $x\in (I\times J)^n$. The partition ${\cal Q} =
 {\cal Q}_1 \wedge {\cal Q}_2$ defining the common knowledge is the
 same as ${\cal Q}_2$ the partition corresponding to the second player.
 There is a one-to-one correspondence between every $C\in {\cal Q}$
  and every sequence $(i_1, j_2, \dots , i_n, j_n)$ of moves by both players. 
  \vskip.2cm

\noi   Let  $S_1$ and $S_2$ be the 
 set of pure decision functions of Player One and Player Two respectively. 
 The space of mixed strategies of
  the truncated game tree is $\Delta:= \Delta (S_1) \times \Delta (S_2)$.
  Likewise a pair of behaviour strategies for the whole game is equivalent to a point in $\Delta$ followed by a collection behaviour strategies for the stages after the $n$th stage.
  Every choice of  $(\sigma, \tau)\in \Delta$ combined with 
 a sequence $i_1, j_1, \dots , i_l, j_l$ of actions taken with 
 positive probability  induces through
 the Bayes rule a conditional probability on $C$. As stated above,
 the sequence 
   $i_1, j_1, \dots , i_l, j_l$ defines uniquely  a member  $C$ in  
 ${\cal Q}$ and $P_{\sigma,\tau} (\, \cdot \,| C)$ is that conditional
 probability, whereby it does not matter whether we
 see this as a distribution on the set
 $C=\{ (k, i_1, j_1, \dots, i_n, j_n) \ | \ k\in K\}$ or on the set
  $K$ itself. \vskip.2cm 

\noi 
Notice that for  $e=(k, i_1, j_1, \dots, i_n, j_n)$,  the probability $P_{\sigma,\tau} (e)$ is the product of 
the pobability of the choice of $k\in K$, 
the corresponding probabilities of actions of  Player One  induced by $\sigma$  and 
the corresponding probabilities of actions of  Player Two induced by $\tau$. 
Since the  probabilities of actions of  Player Two induced by $\tau$ do not depend on %of 
$ k\in K$ we obtain that 
 \vskip.3cm 

{\noi ($\ast$)} If for $\sigma \in \Delta (S_1)$ and $\tau, \tau' \in \Delta (S_2)$ and some $C \in {\cal Q}$ 
both $P_{\sigma,\tau} (C)$ and $P_{\sigma,\tau'} (C)$ are non-zero then the conditional probabilities 
$P_{\sigma,\tau} (\, \cdot \, |C)$ and $P_{\sigma,\tau'} (\, \cdot \, |C)$ are equal. 
 
 \vskip.3cm 

 \noi We define $F_C:\Delta(C) \rightarrow {\bf R}^{C\times \{ 1,2\}}$
   such that for every $p\in \Delta (C)$ the set
   $F_C(p)$ is the convexification of the joint plan equilibria
    corresponding to the initial probability distribution $\Delta (C)$. 
    For every $e\in E$, which corresponds to a history  $h=(k,i_1, i_2, \dots ,
    i_n, j_n)\in {\cal H}_n$, also a member of some $C\in {\cal Q}$, and some
     continuation vector $v\in {\bf R}^{C\times \{ 1,2\}}$, define the payoff 
       $g^{e,i}(v^{e,i})$ to be  $f^i(h)+ (1-\lambda^i) v^{e,i}$. 

     \vskip.2cm
     \noi Now consider the case of $(\sigma, \tau)$ such that the conditional probability on some $C$ is ill defined. If
      there is no $\overline {\tau}$ such that with
      $( \sigma ,\overline {\tau })$ the set $C$ is reached with positive probability,
      then a continuation payoff  can be  chosen arbitrarily in $F_C(q)$ for any $q$. If there is
      some $\overline {\tau} $ such that the set $C$ is reached with positive probability
      with $( \sigma ,\overline {\tau})$, let the continuation
      payoff be any in $F_C(q)$ for $q$ being
       the conditional probability defined by $(\sigma, \overline {\tau})$. 
       Notice that, by ($\ast$), all such $\overline {\tau} $ define the same conditional
      probability. 

     \vskip.2cm 
    \noi To apply Theorem 3, we need to know that $F_C$ so defined
     is u.s.c., non-empty, and convex valued. 
 With regard to the conventional infinitely repeated undiscounted  games,  
   by  Simon, Spie{\.z}, and Toru{\'n}czyk (1995)
  joint plan equilibria exist for every probability in the probability 
 simplex $\Delta (K)$ and the equality and inequality 
  conditions defining them imply that they 
 are upper-semi-continuous as a correspondence (indeed satisfying the %stronger 
more general condition of ``spanning'', Simon, Spie{\.z}, and Toru{\'n}czyk (2002)). 
    It  follows from Hart (1985) that equilibrium payoffs 
 are generated by    convexifying   the payoffs from joint plan 
equilibria corresponding to any fixed probability $p\in \Delta (K)$. 
 As the vector space of payoffs is finite dimensional, 
  the point-wise convexification of an upper-semi-continuous correspondence 
  is also upper-semi-continuous. \vskip.2cm

  \noi From Kuhn's Theorem (1953) we can equivalently consider
  mixed strategies for the first $n$ stages combined with
   behaviour strategies for the following stages. From Theorem 3, there are 
  mixed strategies $\sigma$ and $\tau$ in $\Delta = \Delta (S_1) \times
  \Delta (S_2)$ on the first $n$ stages that satisfy the results  of Theorem 3. 
  We combine the $\sigma$ and $\tau$ with behaviour strategies
   for the remaining stages that correspond, for each $C\in {\cal Q}$, 
   to the equilibrium payoffs in  $F_C (p)$   obtained from 
    Theorem 3. \vskip.2cm 
    \noi In the definition of $\sigma$ and $\tau$, as long as the set $C\in {\cal Q}$ should be  reached with positive probability by
    these strategies,  neither  player cannot  detect   deviation by the other player. Furthermore no action of either player
    in the first $n$ stages can change the conditional probability on any $C\in {\cal Q}$. This is because the actions taken
    by both players define the $C\in {\cal Q}$ and the only way to update the   conditional probabilities 
    is through observation of the played actions. Changing strategies can only result in a change in the distribution
     on the $C\in {\cal Q}$ reached, but not the conditional probability associated with any fixed $C\in {\cal Q}$. \vskip.2cm

    \noi   We consider first what happens at the stages beyond the $n$th, and consider
     first the payoff of Player Two.
  Because of the way 
  the continuation payoff  was defined in all cases  and because the first player
   is adhering to its prescribed  strategy, it does not matter
   whether or not the 
   $C\in {\cal Q}$ is  reached 
   with positive probability, the $q$ used to define the  continuation
   payoff in $F_C(q)$ is the conditional probability on the states as
    defined by the first $n$ stages of play. 
    As his  prescribed behaviour after the $n$th stage   is an equilibrium of the 
    the undiscounted Aumann-Maschler game whose distribution
    on the states of nature is that conditional probability  $q$, there is
    no advantage for deviation.
     As for Player One, 
     it doesn't matter which state is chosen
     and what is the corresponding conditional  probability $q$ on the states (as understood by Player Two),
     Player One  gets the corresponding  continuation payoff with the prescribed 
     behaviour strategy and
     according to Blackwell (1956) cannot obtain
     a better payoff no matter which state was chosen. 
\vskip.2cm 

\noi The equalities and inequalities defining the myopic equilibrium,
combined with the lack of incentive to deviate after the first $n$ stages,
 removes any    
 incentive for either player  to deviate   in the first $n$ stages. \hfill $\Box$

  \vskip.2cm 
\noi  
   Dropping the condition of perfect monitoring, we suspect that a proof of 
 equilibrium existence is straightforward as long
 as Player One  has the ability 
 to send distinct non-revealing signals, the same 
  sufficient condition  for equilibria 
 described in  
 Simon, Spie{\.z}, and Toru{\'n}czyk (2002). \vskip.2cm  

 \noi
  For the application of Theorem 3, it is not
 necessary that the payoffs from the initial $n$ stages are related
 in any way to the payoffs from the following undiscounted game. The  only relevance of the first $n$ stages
  to the following stages  is the induced
  probability distribution on the states $K$. We could therefore
  introduce two sets of payoffs, one for a discounted game with
  infinite sequences $\lambda^i_1, \lambda^i_2, \dots$ for
   both players $i=1,2$ and another set of payoffs 
   for an undiscounted game. 
   Arbitrary pairs of payoffs so
    combined together would allow for  $\epsilon$-equilibria
  for every $\epsilon>0$ (by defining the truncated game tree
   from arbitrarily many initial
   stages). 
   But what of $0$-equilibria?
   The obtaining of good payoffs in one of the two games, either the
    undiscounted  or discounted, 
 would be a distraction for obtaining good payoffs in the other game. 
 Even when the  
 payoff matrices for the undiscounted and discounted evaluates
  are the same (as with the  Neyman games),
  the performing  of  joint lotteries  to  convexify the payoffs
  would be a distraction   from the process of playing 
  the discounted game. Therefore to demonstrate an equilibrium here  
    would require   an extension  of Theorem 2  to   
    the ``spanning property'' of Simon, Spie{\.z}, and Toru{\'n}czyk (2002) rather
     than the much simpler property of convex valued. 
   Nevertheless  one would have 
   to show also that  the equilibrium  behaviour of the players from an
   infinite sequence of 
   game tree truncations would be appropriate for the  undiscounted game. At present we do
   not know if it is possible to
    obtain such an  extension of equilibrium existence to the Neyman games where  there is infinite discounting. 
\vskip.2cm 
\noi Though the theorem can deliver powerful  results concerning the equilibria of composite  games, one 
 does  have to be careful that the given  continuation payoffs  are supported by   equilibria of  the  continuation game. 
  Infinitely repeated games of incomplete information can lack equilibria
 if  one gives to Player Two some very  slight information that Player One 
 does not have; such are  games of ``incomplete information on one and a half sides'' in Sorin and Zamir (1985). Exactly this problem arises because 
 the continuation payoffs of the theorem are determined by a  distribution on the set $C$ that is common knowledge, and yet a player may know 
   more than this and choose not to accept any payoff scheme determined by such common knowledge. In the application to Neyman's question, this problem was avoided by 
 an established theory concerning the equilibria of  games with incomplete information on one side.   
 Indeed even  with  imperfect monitoring over a finite set of stages, there may be problems with 
 the ``individual rationality'' condition necessary for an equilibrium in some contexts (Stapenhorst (2016)).   The desire  not to let such difficulties  
 detract from  the power of the theorem was furthermore  a 
 reason for formulating the theorem without there being necessarily a continuation  game.  

%\vskip2cm
\section{Other examples and an application} %applications  
\noi Although it was developed for understanding the Nash equilibria
of infinitely repeated games, the concept of a myopic equilibrium is
independent of these  games.  \vskip.2cm 

\noi  Look at the following simple example, based on the  $2\times 2$ 
 matrix $ A= \begin{pmatrix} 1 & -1 \cr -1 & 1 \end{pmatrix}$ 
 and   representing 
 the conventional zero-sum 
 matching pennies game with two players and two  actions. From 
 this simple game 
  create  a non-zero-sum game in 
 the following way. If $(p, 1-p)$ is the mixed strategy of Player One and 
 $(q,1-q)$ is the mixed strategy of Player Two (probabilistic choices 
 for the two pure actions), let the payoff of 
 Player One be $(p,1-p) A (q,1-q)^t + \max(p, 1-p)$ and let the payoff 
 of Player Two be $-(p,1-p) A (q,1-q)^t + \max (q, 1-q)$. It is easy 
 to see that there would be no Nash equilibrium in the usual sense of best 
 replies,
 as in response to 
 any mixed strategy of the other player a payoff of $1$ could be obtained 
 by choosing with certainty one or the other    action, and yet 
 a payoff of $1$ could not be obtained by both players simultaneously (as 
  the sum of their payoffs  being at least  
  $2$ is possible only if both 
 chose some action  with certainty and then one of the players 
 would have   a payoff of no more than $0$).  One can also 
 show that this game does not have local equilibria as described above.      
%\noi  Furthermore it is not difficult to show that there is no  
% pair $(p_0, 1-p_0)$ and $(q_0,1-q_0)$
% such that   the payoff for Player One  
% is either myopicly 
% maximal or critical at $p_0$   as a function of $p$  and the 
% same holds for the payoff to Player Two and $q_0$.  

% Clearly this 
% cannot hold if either player is choosing $(\frac 12, \frac 12)$, as 
% a small increase by $w$ in one or the other action (or possibly both) will 
% increase the payoff by at least $w$. 
% If Player One is choosing $(p, 1-p)$ with $0< p < \frac 12$ then to 
% be indifferent between a very  small increase or
% decrease in $p$ it is necessary 
% for $q$ to be $\frac 14$. But then f
  \vskip.2cm \noi

 \noi Now 
 define the payoff from an action $i=1,2$ of Player One as 
 $e_i A (q,1-q)^t + \max (p,1-p)$, where $e_1= (1,0)$ and $e_2= (0,1)$.
 Do the same for Player Two: his payoff is 
$- (p,1-p) A e_i^t + \max (q, 1-q)$.  
Given that both $p$ and $q$ are fixed at $\frac 12$,
 both actions of both players  yield the same expected payoff of $\frac 12$,
 meaning that a myopic equilibrium is defined.
   One could interpret the      
 $(\frac 12, \frac 12)$ distribution   as the  accidental result of 
a flip of the coin  that does not change  
  the probability by which that choice is made.
\vskip.2cm 

\noi To demonstrate  further 
 the fundamental  difference in equilibrium concepts, 
  look at the following  one-person optimisation example where 
 there is both a Nash equilibrium and a myopic equilibrium, but they are 
 very different.
 Our single player  Piers
  wants to vote for Donald Trump, but is
deeply  embarrassed by the desire to do so.
 Behaviour in the voting booth  is secret, however 
 the voting intention of Piers  before entering 
 the voting booth  is not secret (at least from his wife and closest friends)  and this 
influences the utility of his behaviour. 
 Let us assume that $p$ is the probability that Piers will  vote for Trump 
 and that  
 Piers  loses $5p$ in utility through that voting intention, 
 regardless of what he actually does.  
  All things 
 being equal, regardless of the value of $p$, in 
 the voting booth  there is  
 an advantage of $1$ to vote for Trump over Clinton.
 Without loss of generality, lets assume 
 that once in the voting booth the 
  utility of  
voting for  Trump and Clinton
 is    $1-5p$ and  $-5p$ respectively. Regardless of the probability $p$, 
 voting  for Trump is  always  
 preferable to voting for Clinton,
 which makes for one unique myopic equilibrium,  
  namely 
 a certain vote for Trump ($p=1$). 
Define the  payoff function on the probability simplex in 
 the way outlined above --  
 as a function of $p$, the expected utility to Piers
  would be 
 $p(1-p) + (1-p) (-5p) = -4p$.   
The  unique  optimal  payoff  as a function of $p$  would be  $0$  
 obtained at $p=0$, meaning a certain vote for Clinton (and 
 this  defines the unique Nash equilibrium). However the certain 
 vote for Trump, the unique myopic equilibrium,  results in an expected 
  payoff of $-4$. We see from this example  that a myopic equilibrium of a one-player game  is not 
  necessarily a local  maximum. \vskip.2cm \noi

\noi The distinction between myopic equilibria and Nash equilibria for 
 one player games 
 can exist when neither occur at the boundary of the probability simplex. 
   Now we assume, for whatever reason,  that the  embarrassment
 of wanting to vote for Trump  disappears 
 when one actually votes for Clinton. 
 Following this idea, 
  the utility for voting for Trump and Clinton could be  
 $1-5p$ and $0$, respectively.
  As 
 a problem of optimisation, the expected utility of
 the distribution $(p, 1-p)$ is $(1-5p) p= p-5p^2$, a strictly 
 concave function 
 with a unique maximal solution.  
 By taking the derivative and setting it to zero, one 
 discovers that the  value  is maximised at $p=\frac 1{10}$ (the unique myopic  
 maximum and Nash equilibrium)  for 
 the value of $\frac 1 {20}$. 
 However the unique myopic   equilibrium is obtained at $p= \frac 15$, where 
  both the utility of voting for Clinton and voting for Trump are equal and 
 are equal to  
 $0$.
 \vskip.2cm
 \noi   We believe  the most relevant application of myopic equilibria will be toward
 a new and more liberal understanding of what is a subgame in a game tree.
 Conventionally, the concept of a subgame is very restrictive; it is a node where upon
 being reached all players know that this node and only this node has been reached.
 It is common for students  to identity  subgames erroneously because of this
 restrictive definition.
 With the concept of myopic equilibria, for a subset of nodes intermediate to the flow of
  the game 
  we can perceive a family of  subgames as  distributions
  on this set,  determined by the mixed strategies of players who had acted previously. 
 From %the Structure Theorem of Kohlberg and Mertens, 
 Theorem 1 we know
  that the equilibrium correspondence as a function of these distributions has a topological
  structure implying the spanning property of Simon, Spie{\.z}, and Toru{\'n}czyk (2002). This orientation
  would be empowered by a generalisation of Theorem 2 employing the spanning property, both in
  the resulting structure of myopic equilibrium solutions and in the input correspondence
   of payoffs.

\section  {References}

\begin{description}

\item Aumann, R. and Maschler, M. (1995),  
{\it Repeated Games with Incomplete 
Information}. 
 With the collaboration of R. Stearns. 
Cambridge, MA:
M.I.T. Press.

\medskip 

\item Biasi, C. and Monis, T. (2013), Weak Local Nash Equilibrium, 
 {\it Topological  Methods of Nonlinear Analysis,} {\bf 41}, 
 No. 2, pp. 409-419. 
\medskip 

\item Blackwell, D. (1956), An Analogue of the Minimax Theorem for Vector 
 Payoffs, {\em Pacific Journal of Mathematics}, {\bf 6}, pp 1-8. \medskip 

\item Hart, S. (1985),  
Non-zero Sum Two-Person Repeated Games with Incomplete 
 Information, {\it Mathematics of Operations Research} {\bf 10}, No. 1, 
 117-153.
 
\item \red{Held, M., Wolfe, P., Crowder, H. (1974), Validation of subgradient optimisation, {\it Mathematical Programming} {\bf 6}, 62-88.} 

\medskip 
\item Kohlberg, E. and Mertens, J.-F. (1986),  
On the Strategic Stability
 of Equilibria, {\it Econometrica}, {\bf 54 (5)},  pp. 1003-1037.

\medskip 
 \item Kuhn, H. (1953),  Extensive Games and the Problem 
 of Information, in {\em Contributions to the Theory of Games I}, Princeton 
 University Press, eds. 
 Kuhn and Tucker, pp. 193-216.
\medskip 
\item Nash, J. (1950),   Equilibrium Points in $n$-Person Games, 
Proceedings of the National Academy of Sciences, {\bf 36}, pp. 48-49.
\medskip
 \item Neyman, A. (2016), private communication. 
 \medskip 
\item Selten, R. (1975), A Reexamination of the Perfectness Concept for 
 Equilibrium Points in Extensive Games, {\it International 
 Journal of Game Theory}, Vol 4, No. 1, pp 25-55.

 \item Simon, R.S., Spie{\.z}, S., and Toru{\'n}czyk, H. (1995), 
 The Existence of Equilibria in Certain Games,
 Separation for Families of Convex
   Functions and a Theorem of Borsuk-Ulam Type,
   {\it Israel Journal of
    Mathematics}, Vol 92, pp. 1-21.
\medskip

\item   Simon, R.S., Spie{\.z}, S., Toru{\'n}czyk, H. (2002), 
Equilibrium Existence and Topology in Games of Incomplete 
Information on
 One Side,   {\it Transactions
 of the American Mathematical Society}, Vol. 354, No. 12, 
 pp. 5005-5026. \medskip 
\item 
Sorin, S. and Zamir, S.   
 (1985), A Two-Person 
 Game with Lack of Information 
 on One and One-Half          Sides, {\it Mathematics of Operations Research} 
  {\bf 10},  17-23.

\item Stapenhorst, C. (2016), Noisy Signalling in the Principal-Agent Problem, 
 M. Sc. Dissertation, Mathematics, L.S.E. 
\end{description}

\end{document}